\documentstyle[preprint,aps,epsf,floats,amssymb]{revtex}
\tighten
\draft
\widetext
\def\beq{\begin{equation}}\def\eeq{\end{equation}}
\def\beqa{\begin{eqnarray}}\def\eeqa{\end{eqnarray}}
\newcommand{\upr}{\mbox{$U(1)'$}}
\newcommand{\zpr}{\mbox{$Z'$}}
\def\mxth{\mathsurround=0pt }
\def\xversim#1#2{\lower2.pt\vbox{\baselineskip0pt \lineskip-.5pt
  \ialign{$\mxth#1\hfil##\hfil$\crcr#2\crcr\sim\crcr}}}             
\def\simgr{\mathrel{\mathpalette\xversim >}}
\def\simle{\mathrel{\mathpalette\xversim <}}

\newcommand{\drawsquare}[2]{\hbox{%
\rule{#2pt}{#1pt}\hskip-#2pt%  left vertical
\rule{#1pt}{#2pt}\hskip-#1pt%  lower horizontal
\rule[#1pt]{#1pt}{#2pt}}\rule[#1pt]{#2pt}{#2pt}\hskip-#2pt%  upper horizontal
\rule{#2pt}{#1pt}}% right vertical

\newcommand{\fund}{\raisebox{-.5pt}{\drawsquare{6.5}{0.4}}}%  fund
\newcommand{\antifund}{\overline{\fund}}

\preprint{UPR-0990T, NSF-ITP-02-38, MADPH-02-1267, NI02010-MTH}
\bigskip
\bigskip
\begin{document}
\title{Phenomenology of A Three-Family 
%Supersymmetric
Standard-like String Model}
\medskip
\author{Mirjam Cveti\v c$^{1,3,5}$, Paul Langacker$^{1,2,3,4}$, and
Gary Shiu$^{1,3}$ \vspace{0.2cm}}
\address{$^1$ Department of Physics and
Astronomy, University of Pennsylvania, Philadelphia, PA 19104 \\
$^2$ School of Natural Sciences, Institute for Advanced Study,
Princeton, NJ 08540 \\
$^3$ Institute for Theoretical Physics, University of California,
Santa Barbara, CA 93106 \\
$^4$  Department of Physics, University of Wisconsin, Madison, WI 53706\\
$^5$ Isaac Newton Institute for Mathematical Sciences, University of Cambridge,
Cambridge.U.K.}
%\date{January 15, 2002}
\bigskip
\medskip
\maketitle

%%%%%%%%%%%%%%%%%%%%%%%%%%%%%%%%%%%%%%%%%%%%%%%%%%%%%%%%%%%%%%%%%%%%%%%
% Macro
\def\kl#1{\left(#1\right)}
\def\th#1#2{\vartheta\bigl[{\textstyle{  #1 \atop #2}} \bigr] }
%%%%%%%%%%%%%%%%%%%%%%%%%%%%%%%%%%%%%%%%%%%%%%%%%%%%%%%%%%%%%%%%%%%%%%%

\begin{abstract}
{}We discuss the phenomenology of a
three-family supersymmetric Standard-like Model derived from the
orientifold construction,
in which the ordinary chiral states are localized
at the intersection of branes at angles.
In addition to the Standard Model group, there are two
additional \upr \ symmetries, one of which has family non-universal 
and therefore flavor changing couplings, and a quasi-hidden non-abelian
sector which becomes strongly coupled above the electroweak scale. 
The perturbative spectrum contains a fourth family of exotic ($SU(2)$-
singlet) quarks and leptons, in which, however, the left-chiral states
have unphysical electric charges. It is argued that these decouple
from the low energy spectrum due to hidden sector charge confinement, 
and that anomaly matching requires the physical left-chiral states
to be composites. The model has multiple Higgs doublets and
additional exotic states. The moduli-dependent predictions 
for the gauge couplings are discussed.
The strong coupling agrees with experiment for reasonable moduli,
but the electroweak couplings are too small.

\end{abstract}

\section{Introduction}

Despite an enormous amount of promise and interest in superstring/M theory, it
is still difficult to construct models  with a fully realistic effective low energy
four-dimensional limit from first principles. The (related) difficulties include the fact
that only certain limiting corners of M theory are perturbative, there are
many possible compactifications of the extra dimensions, the cosmological
constant, the method of supersymmetry breaking, the difficulty of constructing a
realistic spectrum for the effective theory, and the stabilization of moduli.
One direction, the subject of this paper, is to examine concrete semi-realistic 
constructions in as much detail as possible. One does not expect to find
a fully realistic model. Rather, the goals are: (1) to develop calculational 
techniques,
 (2) to suggest promising
directions for new constructions, and (3)
to find examples of possible new physics at the TeV scale that might
emerge from explicit top-down string constructions.
The latter are sometimes different from new
physics motivated from bottom-up approaches. Of course, the features of a
particular model may simply be shortcomings of that construction rather than
generic predictions of classes of string theories. However, by examining promising
constructions from a variety of corners of M theory one may obtain hints as to
likely types of new TeV scale physics. Difficulties of this program are that most
constructions are rather complicated, and that  the unrealistic or unsuccessful
aspects of each model often make it difficult to carry out detailed calculations.

For over a decade,
there has been considerable effort in the construction of semi-realistic
string models in the framework of perturbative 
heterotic string theory\cite{review}.
In particular, a class of free-fermionic string models which contain
the gauge group and matter content of the
minimal supersymmetric standard model (MSSM) have been
constructed~\cite{fny,CHL}.  These constructions have many
interesting features, such as extended gauge structures and
matter content~\cite{chl5}. Some of the features of one of these
constructions summarized in the Appendix.

The purpose of this paper is to examine the phenomenological issues
in another calculable regime of M theory, namely, Type II orientifolds.
In recent years, the advent of D-branes 
has facilitated the
construction of semi-realistic
string models using conformal field theory techniques, 
as illustrated by the various
four-dimensional ${\cal N}=1$ supersymmetric Type II orientifolds 
($\!$\cite{ABPSS,berkooz,N1orientifolds,ShiuTye,KST,afiv,wlm,CPW,kr,CUW,CSU,CSU-G2} 
and references therein).
%%Changes here
A promising direction to obtain chiral theories is by constructing models
with D-branes
intersecting at angles \cite{bdl}. This fact (or its T-dual version, i.e.,
branes with flux) has been exploited in \cite{bgkl,afiru,bkl} to
construct semi-realistic string models (see also \cite{Angelantonj}).
%%Changes end
However, the constraints on supersymmetric four-dimensional models are
rather restrictive.
Despite the remarkable progress in developing techniques
of orientifold constructions, 
there is only
one orientifold 
model \cite{CSU,CSU-G2} that has been constructed so far 
with the ingredients of the MSSM:
${\cal N}=1$ supersymmetry, the
Standard Model (SM) gauge group as a part of the gauge structure, and candidate
fields for the three generations of quarks and leptons as well as the
electroweak Higgs doublets.
We hope that by studying the phenomenology of this model
in detail, we can probe
some of the generic features and predictions of string
models derived from the orientifold approach.

In this paper we concentrate on direct compactifications of the underlying M
theory to a four-dimensional field theory containing the MSSM (i.e., without
having an intermediate four-dimensional grand unified theory). We  focus 
on the case in which the fundamental scale is comparable
to the Planck scale, i.e., the case with  no very large extra dimensions.

In Section~\ref{description} we briefly summarize the construction of the
model, including the gauge factors and the quantum numbers of
the chiral and non-chiral spectrum. The properties of the
perturbative spectrum are discussed in more detail in Section~\ref{perturbative},
including the properties of the multiple Higgs doublets, the three regular
families, the fourth exotic family, and alternative assignments.
The properties of the additional \upr \ gauge interactions
and the possibilities for breaking them at the electroweak
or intermediate scales are discussed in Section~\ref{additional}.
Section~\ref{gauge} is concerned with the gauge couplings. The model does
not have the conventional form of gauge unification because each  group factor is
associated with a different set of branes. However, the string-scale couplings are
predicted in terms of the ratio of the Planck to string scales and a
geometric factor.
 The low energy electroweak couplings are too small due to
 the multiple Higgs fields and exotic matter, while
 the strong coupling is more reasonable.  The quasi-hidden
 sector groups are asymptotically free.
The implications
of these results for the spectrum are described in Section~\ref{strongly}.
In particular, the fractionally charged exotic states presumably
disappear from the low energy spectrum due to hidden sector charge
confinement, to be replaced by composite states with the appropriate
quantum numbers to form the left-handed components of an exotic
fourth family. The results are summarized and contrasted with
a particular heterotic construction in
Section~\ref{discussion}. A more detailed description of the features
of that construction is given in the Appendix.
A detailed discussion of the Yukawa
couplings of the model will be presented separately~\cite{yukawa},
and further implications of the strong couplings for
moduli stabilization and supersymmetry breaking is under
investigation~\cite{strongsector}.

\section{Description of the Model}
\label{description}

For completeness, let us describe the construction of the model in \cite{CSU},
which is obtained by compactifying Type IIA string theory on a 
${\bf T}^6/({\bf Z}_2\times {\bf Z}_2)$ orientifold.
We considered
a general framework in which the
D-branes are not necessarily parallel to
the orientifold planes, which gives rise to new possibilities
of embedding the gauge sector in the same
background geometry. 
(In the T-dual picture \cite{berkooz},
this corresponds to turning 
on a background flux of the gauge fields on the D-branes).

The generators $\theta$, $\omega$ of ${\bf Z}_2 \times {\bf Z}_2$
act as $ \theta:   (z_1,z_2,z_3) \to (-z_1,-z_2,z_3)$, and  $\omega:
(z_1,z_2,z_3) \to (z_1,-z_2,-z_3)$ on the complex coordinates $z_i$ of 
${\bf T}^6={\bf T}^2 \times {\bf T}^2 \times {\bf T}^2$.
The orientifold
action  is $\Omega R$, where $\Omega$ is world-sheet parity, and $R$ acts
by $ R:\  (z_1,z_2,z_3) \to (\overline{z}_1,\overline{z}_2,\overline{z}_3)$. 
The model
contains four kinds of O6-planes, associated with the actions of
$\Omega R$, $\Omega R\theta$, $\Omega R \omega$, $\Omega R\theta\omega$.
The closed string sector
contains gravitational supermultiplets as well as orbifold moduli
and is straightforward to determine, and so in what follows we will focus
on the open string (charges) spectrum. The cancellation of the RR 
tadpoles from the orientifold planes 
requires an introduction of $K$ stacks of $N_a$ D6-branes
($a=1,\ldots, K$) wrapped on three-cycles $[\Pi_a]$ 
(which for simplicity is taken to be the product of 1-cycles of the two-tori):
\begin{equation}
[\Pi_a] = \prod_{i=1}^3 \,(n_a^i\, [a_i]\, +\, m_a^i\, [b_i])
\end{equation}
and similarly for their orientifold images under
$\Omega R$.
The cycles that the D6-branes and their orientifold images wrap around are
specified by the wrapping numbers $(n_a^i,m_a^i)$. The number of D6-branes
and the associated wrapping numbers are constrained by (i) cancellation
of RR tadpoles, and (ii) supersymmetry.

Analogous to the situation in \cite{bkl},
models with all tori orthogonal lead to an even number of
families. 
%%Changes here
Hence, as in \cite{bkl},
%%Changes end
we consider models with one tilted $T^2$, where the
tilting parameter is discrete and has a unique non-trivial value \cite{ab}. 
This mildly modifies the closed string sector, but has an
important impact on the open string sector. Namely, a D-brane 1-cycle
$(n_a^i,m_a^i)$ along a tilted  torus is mapped to $(n^i,-m^i-n^i)$. It is
convenient to define ${\widetilde m}^i=m^i+\frac 12 n^i$, and label the
cycles as $(n^i,{\widetilde m}^i)$. 

Let us summarize the results for D6-branes not parallel
to O6-planes (for zero angles, the spectrum follows from \cite{berkooz}).
The $aa$ 
sector (strings stretched within a single stack of D6$_a$-branes) is
invariant under $\theta$, $\omega$, and is exchanged with $a'a'$ by the
action of $\Omega R$. For the gauge group, the $\theta$ projection breaks
$U(N_a)$ to $U(N_a/2)\times U(N_a/2)$, and $\omega$ identifies both
factors, leaving $U(N_a/2)$. Concerning the matter multiplets, we
obtain three adjoint $N=1$ chiral multiplets. 

The $ab+ba$ sector (strings stretched between D6$_a$- and D6$_b$-branes)
is invariant, as a whole, under the orbifold projections, and is mapped to
the $b'a'+a'b'$ sector by $\Omega R$. The matter content before any
projection would be given by $I_{ab}$ chiral fermions in the bifundamental
$(\fund_{a},\antifund_b)$ of $U(N_a) \times U(N_b)$, where 
\begin{eqnarray}
I_{ab} = 
%\prod_{i=1}^3 \left( n_a^i m_b^i - n_b^i m_a^i \right) = 
\left( n_a^1 m_b^1 - n_b^1 m_a^1 \right)  
\left( n_a^2 m_b^2 - n_b^2 m_a^2 \right) 
\left( n_a^3 {\widetilde m}_b^3 - n_b^3 {\widetilde m}_a^3 \right) 
\equiv \prod_{i=1}^{3} I_{ab}^{i} \nonumber
\end{eqnarray}
is the intersection number 
%%Changes here
of the wrapped cycles (see \cite{bgkl,afiru}),
%%Changes end 
and the sign of $I_{ab}$
denotes the chirality of the corresponding fermion ($I_{ab}<0$ gives
left-handed fermions in our convention). For supersymmetric
intersections, additional massless scalars complete the corresponding
chiral supermultiplet. In principle, one needs to take into account the
orbifold action on the intersection point. However the final result
turns out to be insensitive to this subtletly and is still given by 
$I_{ab}$ chiral multiplets in the $(\fund_a,\antifund_b)$ of $U(N_a/2)
\times U(N_b/2)$. A similar effect takes place in the $ab'+b'a$ sector, for
$a\neq b$, where the final matter content is given by $I_{ab'}$ chiral
multiplets in the bifundamental $(\fund_a,\fund_b)$. 

For the $aa'+a'a$ sector the orbifold action on the intersection points
turns out to be crucial. For intersection points invariant under the
orbifold, the orientifold projection leads to a two-index antisymmetric
representation of $U(N_a/2)$, except for states with $\theta$ and $\omega$
eigenvalue $+1$, where it yields a two-index symmetric representation.
For points not fixed under some orbifold element, say two points fixed
under $\omega$ and exchanged by $\theta$, one simply keeps one point, and
does not impose the $\omega$ projection. Equivalently, one considers all
possible eigenvalues for $\omega$, and applies the above rule to read off
whether the symmetric or the antisymmetric survives. A closed formula for
the chiral piece in this sector can be found in \cite{CSU}. 

In addition to the chiral multiplets, there can be vector-like
multiplets from the $ab+ba$, $ab'+b'a$ and $aa'+a'a$ sectors.
This happens when 
$I_{ab}^i=0$ for a single $i=i_0$, the intersection number is zero.
Instead of having only left-handed or right-handed chiral multiplets
at the intersection, both the left-handed and the right-handed
chiral multiplets are present. The multiplicity of vector-like
multiplets is given by
\begin{equation}
I_{ab} = \prod_{i \not= i_0} I_{ab}^i
\end{equation}
These vector-like multiplets are generically massive as explained 
in \cite{CSU}.

The condition that the system of branes preserve $N=1$ supersymmetry
requires \cite{bdl} that each stack of D6-branes  is related to
the O6-planes by a rotation in $SU(3)$: denoting by $\theta_i$ the angles
the D6-brane forms with the horizontal direction in the $i^{th}$
two-torus, supersymmetry preserving configurations must
satisfy
$
\theta_1\, +\, \theta_2\, +\, \theta_3\, =\, 0
$.
In order to simplify the supersymmetry conditions within our
search for realistic models, we considered a particular ansatz:
$(\theta_1,\theta_2,0)$, $(\theta_1,0,\theta_3)$ or $(0,\theta_2,\theta_3)$. 

Due to the smaller number of O6-planes in tilted configurations, the RR
tadpole conditions are very stringent for more than one tilted torus.
Focusing on tilting just the third torus, the search for theories with
$U(3)$ and $U(2)$ gauge factors carried by branes at angles and three
left-handed quarks, turns out to be very constraining, at least within our
ansatz. We have found essentially a unique solution. The D6-brane
configuration with wrapping numbers $(n_a^i,\widetilde{m}_a^i)$ is
given in Table \ref{cycles3family}.

\begin{table} 
%\footnotesize
[htb] \footnotesize
\renewcommand{\arraystretch}{1.25}
\begin{center}
\begin{tabular}{|c||c|l|l|} 
Type & $N_a$ & $(n_a^1,m_a^1) \times
(n_a^2,m_a^2) \times (n_a^3,\widetilde{m}_a^3)$ &  Group \\
\hline
$A_1$ & 8 & $(0,1)\times(0,-1)\times (2,{\widetilde 0})$ &  $Q_{8,8'}$ \\
$A_2$ & 2 & $(1,0) \times(1,0) \times (2,{\widetilde 0})$ &  $Sp(2)_A$ \\
\hline
$B_1$ & 4 & $(1,0) \times (1,-1) \times (1,{\widetilde {3/2}})$ & $SU(2)$  \\
$B_2$ & 2 & $(1,0) \times (0,1) \times (0,{\widetilde {-1}})$ & $Sp(2)_B$ \\
\hline
$C_1$ & 6+2 & $(1,-1) \times (1,0) \times (1,{\widetilde{1/2}})$ i&
 $ SU(3), Q_3,Q_1 $ \\
$C_2$ & 4 & $(0,1) \times (1,0) \times (0,{\widetilde{-1}})$  & $Sp(4)$  \\
\end{tabular}
\end{center}
\caption{\small D6-brane configuration for the three-family model.}
\label{cycles3family}
\end{table}
%\begin{table} 
%%\footnotesize
%[htb] \footnotesize
%\renewcommand{\arraystretch}{1.25}
%\begin{center}
%\begin{tabular}{|c||c|l|} 
%Type & $N_a$ & $(n_a^1,m_a^1) \times
%(n_a^2,m_a^2) \times (n_a^3,\widetilde{m}_a^3)$ \\
%\hline
%$A_1$ & 8 & $(0,1)\times(0,-1)\times (2,{\widetilde 0})$ \\
%$A_2$ & 2 & $(1,0) \times(1,0) \times (2,{\widetilde 0})$ \\
%\hline
%$B_1$ & 4 & $(1,0) \times (1,-1) \times (1,{\widetilde {3/2}})$ \\
%$B_2$ & 2 & $(1,0) \times (0,1) \times (0,{\widetilde {-1}})$ \\
%\hline
%$C_1$ & 6+2 & $(1,-1) \times (1,0) \times (1,{\widetilde{1/2}})$ \\
%$C_2$ & 4 & $(0,1) \times (1,0) \times (0,{\widetilde{-1}})$ \\
%\end{tabular}
%\end{center}
%\caption{\small D6-brane configuration for the three-family model.}
%\label{cycles3family}
%\end{table}
  
The $8$ D6-branes labeled $C_1$ are split in two parallel but not
overlapping stacks of $6$ and $2$ branes, and hence lead to a gauge group
$U(3)\times U(1)$. 
Interestingly, a linear combination of the two $U(1)$'s is actually 
a generator within the $SU(4)$ arising for coincident branes. This ensures 
that this $U(1)$ is automatically non-anomalous and massless (free of
linear couplings to untwisted moduli) \cite{afiru,imr}, and turns out to
be crucial in the appearance of hypercharge in this model.

For convenience we consider the $8$ D6-branes labeled $A_1$ to be away
from the O6-planes in all three complex planes. This leads to
two D6-branes that can move independently (hence give rise to a group
$U(1)^2$), plus their $\theta$, $\omega$ and $\Omega R$ images. 
These $U(1)$'s are also automatically non-anomalous and massless.
In the effective theory, this corresponds to Higgsing of $USp(8)$ down to
$U(1)^2$.

The surviving non-abelian gauge group is
$SU(3)\times SU(2)\times Sp(2)\times Sp(2)\times Sp(4)$.
The $SU(3)\times SU(2)$ corresponds to the MSSM, while the
last three factors form a quasi-hidden sector (i.e., most states
are charged under one sector or the other, but there are
a few which couple to both.) In addition, 
there are three non-anomalous $U(1)$ factors and two anomalous ones.
The
generators $Q_3$, $Q_1$ and $Q_2$ refer to the $U(1)$ factor within the
corresponding $U(n)$, while $Q_8$, $Q_8'$ are the $U(1)$'s arising from
the $USp(8)$. $Q_3/3$ and $Q_1$ are essentially baryon ($B$) and
lepton ($L$) number, respectively, while $(Q_8+Q_8')/2$ is
analogous to the generator $T_{3R}$ occurring in left-right
symmetric extensions of the SM.
The hypercharge is defined as:
\beqa
Q_Y & = & \frac 16 Q_3 - \frac 12 Q_1 + \frac 12 (Q_8+Q_8'). 
\label{hyper}
\eeqa
From the above comments, $Q_Y$ as defined guarantees that $U(1)_Y$ is
massless. There are two additional surviving non-anomalous $U(1)$'s,
i.e., $B-L = Q_3/3-Q_1$ and $Q_8-Q_8'$. The gauge bosons
corresponding to the anomalous $U(1)$ generators $B+L$ and $Q_2$
acquire string-scale masses, so those generators act like 
global symmetries on the effective four-dimensional theory.

The spectrum of chiral multiplets in the open string sector
is tabulated in Table \ref{spectrum3}. 
For completeness, we also give the spectrum of the vector-like multiplets 
in Tables \ref{vectorab}, \ref{vectorab'}
and \ref{vectoraa'} even though these multiplets are generically massive.
The theory contains three Standard Model families, multiple Higgs
candidates, a number of exotic
chiral (but anomaly-free)  fields, and multiplets which transform as
adjoints or singlets under the SM
gauge group.The spectrum is discussed in more detail in the following section.

\begin{table} \footnotesize
%[htb] \footnotesize
\renewcommand{\arraystretch}{1.25}
\begin{center}
\begin{tabular}{|c||c||c|c|c|c|c||c|c|c|}
%Sector & $U(3)\times U(2)\times USp(2)\times USp(2)\times USp(4)$ &
Sector & $SU(3)\times SU(2)\times Sp(2)_B\times Sp(2)_A\times Sp(4)$ &
$Q_3$ & $Q_1$ & $Q_2$ & $Q_8$ & $Q_8'$ & $Q_Y$ & $Q_8-Q_8'$ & Field
\\
\hline
%$A_1 A_1$ & $2\times 3 \times (1,1,1,1,1)$ &
%0 & 0 & 0 & 0 & 0 & 0 & 0 & \\
%$A_2 A_2$ & $3\times (1,1,1,1,1)$ &
%0 & 0 & 0 & 0 & 0 & 0 & 0 & \\
%$B_1 B_1$ & $3 \times (1,3,1,1,1)$ &
%0 & 0 & 0 & 0 & 0 & 0 & 0 & \\
%$B_2 B_2$ & $3 \times (1,1,1,1,1)$ &
%0 & 0 & 0 & 0 & 0 & 0 & 0 & \\
%$C_1 C_1$ & $3 \times (8+1,1,1,1,1)$ &
%0 & 0 & 0 & 0 & 0 & 0 & 0 & \\
%          & $3 \times (1,1,1,1,1)$ &
%0 & 0 & 0 & 0 & 0 & 0 & 0 & \\
%$C_2 C_2$ & $3 \times (1,1,1,1,5+1)$ &
%0 & 0 & 0 & 0 & 0 & 0 & 0 & \\
%\hline
$A_1 B_1$ & $3 \times 2\times (1,{\overline 2},1,1,1)$ &
0 & 0 & $-1$ & $\pm 1$ & 0 & $\pm \frac 12$ & $\pm 1$ & $H_U$, $H_D$\\
          & $3\times 2\times (1,{\overline 2},1,1,1)$ &
0 & 0 & $-1$ & 0 & $\pm 1$ & $\pm \frac 12$ & $\mp 1$ & $H_U$, $H_D$\\
$A_1 C_1$ & $2 \times (\overline{3},1,1,1,1)$ &
%$-1$ & 0 & 0 & $\pm 1$ & 0 & $\frac 13, -\frac 23$ & $1,-1$ & $U$, $D$\\
$-1$ & 0 & 0 & $\pm 1$ & 0 & $\frac 13, -\frac 23$ & $1,-1$ &
$\bar{D}$, $\bar{U}$\\
          & $2 \times (\overline{3},1,1,1,1)$ &
%$-1$ & 0 & 0 & 0 & $\pm 1$ & $\frac 13, -\frac 23$ & $-1,1$ & $U$, $D$\\
$-1$ & 0 & 0 & 0 & $\pm 1$ & $\frac 13, -\frac 23$ & $-1,1$ &
$\bar{D}$, $\bar{U}$\\
          & $2 \times (1,1,1,1,1)$ &
%0 & $-1$ & 0 & $\pm 1$ & 0 & $1,0$ & $1,-1$ & $E$, $\nu_R$\\
0 & $-1$ & 0 & $\pm 1$ & 0 & $1,0$ & $1,-1$ &
$\bar{E}$, $\bar{N}$\\
          & $2 \times (1,1,1,1,1)$ &
%0 & $-1$ & 0 & 0 & $\pm 1$ & $1,0$ & $-1,1$ & $E$, $\nu_R$\\
0 & $-1$ & 0 & 0 & $\pm 1$ & $1,0$ & $-1,1$ &
$\bar{E}$, $\bar{N}$\\
$B_1 C_1$ & $(3,{\overline 2},1,1,1)$ &
1 & 0 & $-1$ & 0 & 0 & $\frac 16$ & 0 & $Q_L$\\
             & $(1,{\overline 2},1,1,1)$ &
0 & 1 & $-1$ & 0 & 0 & $-\frac 12$ & 0 & $L$\\
$B_1 C_2$ & $(1,2,1,1,4)$ &
0 & 0 & $1$ & 0 & 0 & 0 & 0 & \\
$B_2 C_1$ & $(3,1,2,1,1)$ &
1 & 0 & 0 & 0 & 0 & $\frac 16$ & 0 & \\
          & $(1,1,2,1,1)$ &
0 & 1 & 0 & 0 & 0 & $-\frac 12$ & 0 & \\
$B_1 C_1^{\prime}$ & $2\times (3,2,1,1,1)$ &
1 & 0 & 1 & 0 & 0 & $\frac 16$ & 0 & $Q_L$ \\
                   & $2\times (1,2,1,1,1)$ &
0 & 1 & 1 & 0 & 0 & $-\frac 12$ & 0 & $L$ \\
\hline
$B_1 B_1^{\prime}$ & $2\times (1,1,1,1,1)$ &
0 & 0 & $-2$ & 0 & 0 & 0 & 0 &  \\
                   & $2\times (1,3,1,1,1)$ &
0 & 0 & $2$ & 0 & 0 & 0 & 0 & \\
\hline \hline 
$A_1 A_1$ & $3 \times 8 \times (1,1,1,1,1)$ & 0 & 0 & 0 & 0 & 0 & 0 & 0 &  \\
& $3 \times 4 \times (1,1,1,1,1)$ & 0 & 0 & 0 & $\pm 1$ & $\pm 1$ & $\pm 1$
& 0 &  \\
& $3 \times 4 \times (1,1,1,1,1)$ & 0 & 0 & 0 & $\pm 1$ & $\mp 1$ & 0
& $\pm 2$ &  \\
& $3 \times  (1,1,1,1,1)$ & 0 & 0 & 0 & $\pm 2$ & 0 & $\pm 1$ & $\pm 2$ &  \\
& $3 \times  (1,1,1,1,1)$ & 0 & 0 & 0 & 0 & $\pm 2$ & $\pm 1$ & $\mp 2$ &  \\
$A_2 A_2$ & $3 \times (1,1,1,1,1)$ & 0 & 0 & 0 & 0 & 0 & 0 & 0 &  \\
$B_1 B_1$ & $3 \times (1,3,1,1,1)$ & 0 & 0 & 0 & 0 & 0 & 0 & 0 &  \\
 & $3 \times (1,1,1,1,1)$ & 0 & 0 & 0 & 0 & 0 & 0 & 0 &  \\
$B_2 B_2$ & $3 \times (1,1,1,1,1)$ & 0 & 0 & 0 & 0 & 0 & 0 & 0 &  \\
$C_1 C_1$ & $3 \times (8,1,1,1,1)$ & 0 & 0 & 0 & 0 & 0 & $0$ & 0 &  \\
& $3 \times (1,1,1,1,1)$ & 0 & 0 & 0 & 0 & 0 & $0$ & 0 &  \\
$C_2 C_2$ & $3 \times (1,1,1,1,5+1)$ & 0 & 0 & 0 & 0 & 0 & 0 & 0 &  \\
\end{tabular}
\end{center}
\caption{\small The chiral spectrum of the open string sector in the
three-family model. 
To be complete, we also list  (in the bottom part of the table,
below the double horizontal line) the chiral states from the
$aa$ sectors, which are not localized at the intersections.
\label{spectrum3}}
\end{table}

\begin{table} \footnotesize
%[htb] \footnotesize
\renewcommand{\arraystretch}{1.25}
\begin{center}
%\begin{tabular}{|c||c||c|c|c|c|c||c|c|c|}
%Sector & $U(3)\times U(2)\times USp(2)\times USp(2)\times USp(4)$ &
%$Q_3$ & $Q_1$ & $Q_2$ & $Q_8$ & $Q_8'$ & $Q_Y$ & $Q_8-Q_8'$ & Field
\begin{tabular}{|c||c||c|c|c|c|c||c|c|}
Sector & $SU(3)\times SU(2)\times Sp(2)_B\times Sp(2)_A\times Sp(4)$ &
$Q_3$ & $Q_1$ & $Q_2$ & $Q_8$ & $Q_8'$ & $Q_Y$ & $Q_8-Q_8'$ 
\\
\hline
$A_1 A_2$ & $2 \times (1,1,1,2,1)$ + $2 \times (1,1,1,\overline{2},1)$
& 0 & 0 & 0 & $\pm 1$ & 0 & $\pm \frac{1}{2}$ & $\pm 1$  \\
& $2 \times (1,1,1,2,1)$ + $2 \times (1,1,1,\overline{2},1)$ & 0 & 0 & 0 & 0 & $\pm 1$ &$\pm \frac{1}{2}$ & $\mp 1$  \\
$A_1 B_2$ & $2 \times 2 \times (1,1,2,1,1)$ + $2 \times 2 \times (1,1,\overline{2},1,1)$ 
&0  & 0 & 0  & $\pm 1$ & 0 & $\pm \frac{1}{2}$ & $\pm 1$  \\
& $2 \times 2 \times (1,1,2,1,1)$ + $2 \times 2 \times (1,1,\overline{2},1,1)$ 
&0  & 0 & 0  & 0 & $\pm 1$  &$\pm \frac{1}{2}$ & $\mp 1$  \\ 
$A_1 C_2$ & $2 \times 2 \times (1,1,1,1,4)$ + 
$2 \times 2 \times (1,1,1,1,\overline{4})$ & 0 & 0 & 0 & $\pm 1$ & 0 
& $\pm \frac{1}{2}$ & $\pm 1$  \\
& $2 \times 2 \times (1,1,1,1,4)$ + $2 \times 2 \times (1,1,1,1,\overline{4})$ & 0 & 0 & 0 & 0 & $\pm 1$ & $\pm \frac{1}{2}$ & $\mp 1$  \\
$A_2 B_1$ & $3 \times (1,2,1,\overline{2},1)$  & 0 & 0 & 1 & 0 & 0 & 0 &0  \\
& $3 \times (1,\overline{2},1,2,1)$ 
& 0 & 0 & $-1$ & 0 & 0 & 0 &  0  \\
$A_2 B_2$ & $2 \times (1,1,2,\overline{2},1)$ + $2 \times (1,1,\overline{2},2,1)$ & 0 & 0 & 0 & 0 & 0 & 0 & 0  \\
$A_2 C_1$ & $(3,1,1,\overline{2},1)$
 & 1 & 0 & 0 & 0 & 0 & $\frac{1}{6}$ & 0  \\
& $(\overline{3},1,1,2,1)$
 & $-1$ & 0 & 0 & 0 & 0 & $-\frac{1}{6}$ & 0  \\
& $(1,1,1,\overline{2},1)$
 & 0 & 1 & 0 & 0 & 0 & $-\frac{1}{2}$ & 0  \\
& $(1,1,1,2,1)$
 & 0 & $-1$ & 0 & 0 & 0 & $\frac{1}{2}$ & 0  \\
$A_2 C_2$ & $2 \times (1,1,1,2,\overline{4})$ +
$2 \times (1,1,1,\overline{2},4)$ & 0 & 0 & 0 & 0 & 0 & 0 & 0  \\
$B_1 B_2$ & $(1,2,\overline{2},1,1)$ & 0 & 0 & 1 & 0 & 0 & $-\frac{1}{2}$ & 0  \\
 & $(1,\overline{2},2,1,1)$ & 0 & 0 & $-1$ & 0 & 0 & $\frac{1}{2}$ & 0  \\
$B_2 C_2$ & $(1,1,2,1,\overline{4})$ + $(1,1,\overline{2},1,4)$
 & 0 & 0 & 0 & 0 & 0 & 0 & 0  \\
$C_1 C_2$ & $(3,1,1,1,\overline{4})$ & 1 & 0 & 0 & 0 & 0 & $\frac{1}{6}$ & 0  \\
& $(\overline{3},1,1,1,4)$ & $-1$ & 0 & 0 & 0 & 0 & $-\frac{1}{6}$ & 0  \\
& $(1,1,1,1,\overline{4})$ & 0 & 1 & 0 & 0 & 0 & $-\frac{1}{2}$ & 0  \\
& $(1,1,1,1,4)$ & 0 & $-1$ & 0 & 0 & 0 & $\frac{1}{2}$ & 0  \\
\end{tabular}
\end{center}
\caption{\small Non-Chiral Spectrum from the $ab$ sectors in the
three-family model. These states, as well as those in Tables 
 \ref{vectorab'}
and \ref{vectoraa'},
are generically massive.
\label{vectorab}}
\end{table}

\begin{table} \footnotesize
%[htb] \footnotesize
\renewcommand{\arraystretch}{1.25}
\begin{center}
\begin{tabular}{|c||c||c|c|c|c|c||c|c|}
Sector & $SU(3)\times SU(2)\times Sp(2)_B\times Sp(2)_A\times Sp(4)$ &
$Q_3$ & $Q_1$ & $Q_2$ & $Q_8$ & $Q_8'$ & $Q_Y$ & $Q_8-Q_8'$ 
\\
\hline
$A_1 B_2'$ & $2 \times 2 \times (1,1,2,1,1)$ + $2 \times 2 \times (1,1,\overline{2},1,1)$
& 0 & 0 & 0 & $\pm 1$ & 0 & $\pm \frac{1}{2}$ & $\pm1$  \\
& $2 \times 2 \times (1,1,2,1,1)$ + $2 \times 2 \times (1,1,\overline{2},1,1)$
& 0 & 0 & 0 & 0 & $\pm 1$ & $\pm \frac{1}{2}$ & $\mp1$  \\
$A_1 C_2'$ & $2 \times 2 \times (1,1,1,1,4)$ + $2 \times 2 \times (1,1,1,1,\overline{4})$
& 0 & 0 & 0 & $\pm 1$ & 0 & $\pm \frac{1}{2}$ & $\pm 1$  \\
& $2 \times 2 \times (1,1,1,1,4)$ + $2 \times 2 \times (1,1,1,1,\overline{4})$
& 0 & 0 & 0 & 0 & $\pm 1$ & $\pm \frac{1}{2}$ & $\mp 1$  \\
$A_2 B_1'$ & $3 \times (1,2,1,2,1)$ 
& 0 & 0 & 1 & 0 & 0 & 0 & 0  \\
& $3 \times  (1,\overline{2},1,\overline{2},1)$ 
& 0 & 0 & $-1$ & 0 & 0 & 0 & 0  \\
$A_2 B_2'$ & $2 \times  (1,1,2,2,1)$ + $2 \times (1,1,\overline{2},\overline{2},1)$
& 0 & 0 & 0 & 0 & 0 & 0 & 0  \\
$A_2 C_1'$ & $(3,1,1,2,1)$ 
& 1 & 0 & 0 & 0 & 0 & $\frac{1}{6}$ & 0  \\
& $(\overline{3},1,1,\overline{2},1)$ 
& $-1$ & 0 & 0 & 0 & 0 & $-\frac{1}{6}$ & 0  \\
& $(1,1,1,2,1)$ 
& 0 & 1 & 0 & 0 & 0 & $-\frac{1}{2}$ & 0  \\
& $(1,1,1,\overline{2},1)$ 
& 0 & $-1$ & 0 & 0 & 0 & $\frac{1}{2}$ & 0  \\
$A_2 C_2'$ & $2 \times (1,1,1,2,4)$ +$2 \times (1,1,1,\overline{2},\overline{4})$ 
& 0 & 0 & 0 & 0 & 0 & 0 & 0  \\
$B_1 B_2'$ & $(1,2,2,1,1)$
& 0 & 0 & 1 & 0 & 0 & 0 & 0  \\
& $(1,\overline{2},\overline{2},1,1)$
& 0 & 0 & $-1$ & 0 & 0 & 0& 0 \\
$B_2 C_2'$ & $(1,1,2,1,4)$ +$(1,1,\overline{2},1,\overline{4})$
& 0 & 0 & 0 & 0 & 0 & 0 & 0  \\
$C_1 C_2'$ & $(3,1,1,1,4)$ 
& 1 & 0 & 0 & 0 & 0 & $\frac{1}{6}$ & 0  \\
& $(\overline{3},1,1,1,\overline{4})$ 
& $-1$ & 0 & 0 & 0 & 0 & $-\frac{1}{6}$ & 0  \\
& $(1,1,1,1,4)$ 
& 0 & 1 & 0 & 0 & 0 & $-\frac{1}{2}$ & 0  \\
& $(1,1,1,1,\overline{4})$ 
& 0 & $-1$ & 0 & 0 & 0 & $\frac{1}{2}$ & 0  \\
\end{tabular}
\end{center}
\caption{\small Non-Chiral Spectrum from the $ab'$ sectors in the
three-family model. 
\label{vectorab'}}
\end{table}

\begin{table} \footnotesize
%[htb] \footnotesize
\renewcommand{\arraystretch}{1.25}
\begin{center}
\begin{tabular}{|c||c||c|c|c|c|c||c|c|}
Sector & $SU(3)\times SU(2)\times Sp(2)_B\times Sp(2)_A\times Sp(4)$ &
$Q_3$ & $Q_1$ & $Q_2$ & $Q_8$ & $Q_8'$ & $Q_Y$ & $Q_8-Q_8'$ 
\\
\hline
$B_1 B_1'$ & $4 \times (1,1,1,1,1)$ & 0 & 0 & $ \pm 2$ &0  &0 & 0 & 0  \\
$C_1 C_1'$ & $2 \times (3,1,1,1,1)$ & 2 & 0 & 0 &0  &0 & $\frac{1}{3}$ & 0  \\
& $2 \times (\overline{3},1,1,1,1)$ & $-2$ & 0 & 0 &0  &0 & $-\frac{1}{3}$
& 0  \\
\end{tabular}
\end{center}
\caption{\small Non-Chiral Spectrum from the $aa'$ sectors in the
three-family model. 
\label{vectoraa'}}
\end{table} 

\section{The Perturbative Spectrum}
\label{perturbative}
In this section we describe the simplest version of
the perturbative chiral spectrum. The modifications
expected due to strong coupling effects in the quasi-hidden sector
will be described  in Section~\ref{strongly}.
\begin{itemize}
\item The $A_1B_1$ sector contains 24 Higgs doublets, i.e., twelve
$H_U, H_D$ pairs, where $H_U$ ($H_D$) has the appropriate hypercharge
to generate masses for the charge $2/3$ quarks and neutrinos (the charge
$-1/3$ quarks and charged leptons). Six pairs have
$Q_8$ charges, and the other six $Q_{8'}$ charges.
The existence of so many
doublets is the most unrealistic feature of the model, and is a
major cause for the unrealistic predictions for the low energy gauge couplings
in the MSSM sector. Nevertheless, it at least suggests the possibility
of additional Higgs doublets, and the associated phenomenological
consequences, such as a richer Higgs/neutralino/chargino spectrum and
Higgs-mediated flavor changing or CP violation effects. The problem for this
construction is aggravated by the fact that there is no satisfactory 
mechanism to generate effective supersymmetric masses ($\mu$ terms)
for all of the Higgs multiplets. Elementary $\mu$ terms are not
generated in 
the string construction (and are also forbidden by  $Q_2$). 
An effective $\mu$ would be generated by the VEV of the scalar
component of a superfield $S$ if there were a term $S H_U H_D$
in the superpotential~\cite{SY,ewscale}.
However, there are no $SU(2)$-singlet, $Q_Y=0$
states in the chiral spectrum with the gauge quantum numbers to
 play the role of $S$.
In particular, the $\bar N  H_U H_D$ couplings are forbidden by 
$Q_1$ and $Q_2$.
However, there are non-chiral $B_1 B_1'$ states with the appropriate
quantum numbers.
There are couplings between the $A_1 B_1$, $B_1 B_1'$
and $B_1' A_1$ sectors. Since the $B_1' A_1$ sector is the same as $A_1' B_1$,
 and furthermore, $A_1=A_1'$ (because the $A_1$ brane is the same
as its own orientifold image),  in principle,
there are non-zero couplings of the form $(A_1 B_1)(B_1 B_1')(A_1 B_1)$.
Hence, if these non-chiral states are not massive,
they can play the role of the $S$ field. An explicit calculation
of the above coupling from the corresponding string amplitude, however, is
necessary to verify that there are no additional stringy symmetries to
forbid such couplings.

\item The $B_1C_1'$ intersection yields two families of quark and
lepton doublets, while a third is associated with $B_1C_1$. Only the
first two can have Yukawa couplings to the $H_{U,D}$ because of $Q_2$,
so one fermion family will remain massless. (The Yukawa couplings of
the model are discussed in detail in~\cite{yukawa}.) 
The $A_1C_1$ sector contains four families of $SU(2)$-singlet
antiquarks and antileptons, $\bar U, \bar D, \bar E$, and $\bar N$,
two charged under $Q_8$ and two under $Q_{8'}$.
Three of these should be the partners of the quark and lepton
doublets. 

\item The $B_1C_2$ and $B_2C_1$ states  couple to both the
ordinary and quasi-hidden sector gauge groups, and they carry the fractional
electric charges $\pm \frac 12, \frac 16$. In particular, the
$B_2C_1$ states include two $SU(3)$ triplets and two $SU(3)$ singlets. 
All are $SU(2)$ singlets. These
would be the natural partners of the extra $\bar U, \bar D, \bar E$, and $\bar N$
family from $A_1C_1$ except that they have the wrong electric and 
hypercharges\footnote{We have explored the possibility of using
an alternative hypercharge definition, in which $Q_Y$ includes as an
additional term the diagonal generator of the first $Sp(2)$ group.
However, the required $Sp(2)$ breaking cannot occur without breaking
supersymmetry because in this sector there are only two-branes and thus  
there are not enough branes available to split the branes 
in a ${\bf Z}_2 \times {\bf Z}_2$ invariant manner and compatible with the
orientifold projection.
Breaking of this $Sp(2)$ symmetry  by some 
(perhaps non-perturbative) mechanism at the TeV scale might still be
a possibility, but we  concentrate here on the alternative
mechanism described in  Section~\ref{strongly}.}.
It will be argued in Section~\ref{strongly} that these states most likely
disappear from the physical low energy spectrum due to charge confinement for the
strongly coupled hidden sector gauge groups, to be replaced with
composite fields with the appropriate quantum numbers to be the
($SU(2)$-singlet) partners of the fourth family of $\bar U, \bar D, \bar E$, and
$\bar N$.
\item The $B_1B_1'$ and $aa$ sector chiral supermultiplets
 in Table~\ref{spectrum3}
include three $SU(3)$ octets and five $SU(2)$ triplets, as well
as a number of non-abelian singlets with non-zero hypercharge.
The $aa$ states  are not localized at intersections, and there is no known
mechanism to give them large masses. Hence, they contribute
significantly to the running of the gauge couplings. (We will actually
ignore them, hoping that some mass mechanism will be eventually found.)
These kinds of states (adjoints of the non-Abelian gauge symmetry) 
are a generic problem for  orientifold constructions with
intersecting branes;  since each brane  wraps  a (flat) 
supersymmetric three-cycle of
the  six-torus, these three-cycles are {\it not} rigid and thus the adjoint matter
corresponds the moduli associated with the translational invariance of each
three-cycle. 
One possibility to remove these states would be to put branes on curved
(Calabi-Yau) space with rigid three-cycles, 
which is beyond the scope of the original
construction of the model.
The sector also contains a number  of
SM singlets, and
 three $Sp(4)$
5-plets. 
\item One can consider alternative identifications of some of the
multiplets, due to the fact that the lepton doublets $L$ and
the Higgs doublets $H_D$ have the same SM quantum numbers.
For example, one could interpret one or both of the $SU(3)$-singlets
in $B_1C_1'$ to be $H_D'$ states, and some of the
doublets in $A_1B_1$ as leptons. One could give masses to all of the
charged leptons in this way, but not all of the quarks. Also,
if the scalar component of $\bar N$ acquired a VEV, some, but not all, of
the Higgs fields could have effective $\mu$ terms. We do not pursue
this alternative further.

\end{itemize}

\section{The Additional $U(1)$ Factors}
\label{additional}
As described in Section~\ref{description}, there are
 two  surviving non-anomalous $U(1)$'s,
$B-L = Q_3/3-Q_1$ and $Q_8-Q_8'$, in addition to 
hypercharge\footnote{Strictly speaking, the three  independent \upr \
charges
are $Q_8-Q_8'$, $Q_8+Q_8'$, and $B-L$. One can rotate the corresponding
gauge bosons so that they couple to $Q_8-Q_8'$, $Q_Y$, and
$\beta (Q_8+Q_8')- \frac 1 \beta (B-L)$, where
$\beta \equiv g_{8^+}/g_{B-L}$ is the ratio of $Q_8+Q_8'$ and $B-L$
gauge couplings.}.
The $Q_8-Q_8'$ charges are not family universal for the 
antiquarks and antileptons. For example, two of the $\bar U$ 
have $Q_8-Q_8'=-1$ and two have $+1$. When family mixing
is considered, there may be flavor changing
neutral current (FCNC) couplings of the corresponding $Z$ boson
to the $\bar U$ quarks~\cite{fcnc}, provided that the two types of
$\bar U$'s mix with each other. (This will occur provided there
are $H_U$ scalars of both types, i.e., with $Q_8\ne 0$ and with
$Q_{8'} \ne 0$, with nonzero VEVs.) Similar statements
apply to the $\bar D, \bar E$, and
$\bar N$ couplings. Such couplings could lead to decays such
as $B_s \rightarrow \mu^+ \mu^-$ or $\tau \rightarrow \mu e^+ e^-$.
On the other hand, the $B-L$ charges
are family universal (even though one of the three families of $Q$
and $L$ has a different origin than the other two), so there are
no flavor-violating $B-L$ couplings.

Limits on the mass and mixings of extra \zpr \
bosons depend on their couplings, but  typically
 $M_{Z'} > $ 500--800 GeV, and the $Z-Z'$ mixing
angle  is $<$ few $\times 10^{-3}$~\cite{je}.
One possibility is for a \upr \ to be broken at a large
scale intermediate between the TeV and Planck scales. 
This can occur if there are
two or more scalar fields with opposite signs for their \upr \ charges, so that
the symmetry breaking is along a $D$ (and $F$)-flat direction~\cite{intscale}.
This would be desirable for implementing a neutrino seesaw mechanism if
the heavy singlet neutrino carries a nonzero \upr \ charge.
Alternatively, the breaking can occur at the TeV scale or lower~\cite{ewscale}, 
in which case the minimum need not be supersymmetry conserving.
In either case, one expects to first generate masses for and possibly
 mixing
between the two new \zpr \ bosons via the VEVs of SM-singlet
fields that are large compared to the electroweak scale. A small
mixing of these states with  the $Z$ would then be induced by
Higgs doublet VEVs if they carry \upr \ charges\footnote{In a
multi-Higgs model, one can in principle generate the entire 
$Z-Z'$ mass matrix by the VEVs of Higgs doublets. 
(One of the mass eigenvalues vanishes for a single doublet.)
The mixing can vanish for one or more $H_{U,D}$ pairs (along a $D$-flat
direction for more than one pair).
However, the large \zpr$/Z$ mass
ratio would require a very large \zpr \ coupling to the doublets, which is 
not the case here.}.

The only SM-singlet fields which couple to the extra \upr's are: (1)
the two pairs of $\bar N$ states with $B-L=1$ and
$Q_8=-1$ or $Q_8'=-1$. We refer to these as $\bar N(8)$ and $\bar N(8')$,
respectively; (2) the $Q_Y=B-L=0$ and $Q_8-Q_8' = \pm 2$ states in the
(otherwise problematic) $A_1A_1$ sector. The $A_1A_1$ states, if present
in the spectrum, could lead to a $D$-flat direction for which
$Q_8-Q_8'$ (but not $B-L$) is broken at an intermediate scale.
Depending on whether these $A_1A_1$ states are relevant, one could have
either that $Q_8-Q_8'$ is broken at an intermediate scale and $B-L$ at
the TeV scale (by $\bar N(8)$ and $\bar N(8')$); or that both are
broken at the TeV scale by $\bar N(8)$ and $\bar N(8')$. The second 
possibility could lead to significant mixing between the two \zpr \ bosons.
Of course, symmetry breaking induced by the VEVs of the scalar components
of $\bar N(8)$ or
$\bar N(8')$ would lead to mixing between the corresponding gauginos and the
fermionic sterile neutrinos in $\bar N(8)$  and $\bar N(8')$.

The composite $SU(2)$-singlet $N_4$ supermultiplet suggested by the
strong coupling of the $Sp(2)_B$ group (Section~\ref{strongly})
could also play a role in the breaking of the extra \upr's.
This state is only relevant to the effective theory below the
scale at which $Sp(2)_B$ becomes strong, but that could be anywhere from
a few TeV up to very high scales, so it is possible for $B-L$ to be
broken along a $D$-flat direction  at a large scale.
Of course, other dynamical mechanisms, such as fermion condensates,
could possibly be relevant to the breaking of the extra \upr's.

\section{Gauge Couplings} \label{gauge}
The gauge couplings are associated with different stacks of branes
and do not exhibit conventional gauge unification. Nevertheless,
the value of each gauge coupling at the string scale is predicted
in terms of a modulus $\chi$ and the ratio of the Planck to string scales.
The running is strongly affected by the exotic
matter and multiple Higgs fields, leading to low values of the
MSSM sector couplings at low energy. However, the hidden sector
groups are asymptotically free.

The gauge coupling of the gauge field from a stack of
$D6$-branes wrapping a three-cycle
is given by \cite{CSU} 
(the factors of $\sqrt{2 \pi}$ were carefully
worked out in \cite{ShiuTye}):
\begin{equation}
g_{YM}^2 M_P^{(4d)} = \sqrt{2 \pi} M_s \frac{\sqrt{V_6}}{V_3}
\end{equation}
where $V_3$ is the volume of the three-cycle, $V_6$ is the total internal volume,
and $M_s$ is the string scale.
The  four-dimensional Planck scale  is defined as the coefficient of the Einstein term
in the low energy effective action:
\begin{equation}
S_{4d} = (M_P^{(4d)})^2 \int dx^4 \sqrt{g} R + \dots
= \frac{1}{16 \pi G_N} \int dx^4 \sqrt{g} R + \dots
\end{equation}
Since
$G_N^{-1/2}=1.22 \times 10^{19}\ {\rm GeV}$,
we have
%$M_P^{(4d)} \sim  1.7 \times 10^{18} \ {\rm GeV}$.
$M_P^{(4d)} = \frac{1}{4 \sqrt{\pi}} \times G_N^{-1/2}
= 1.7 \times 10^{18}$ GeV.
(Some authors define a scale larger by $\sqrt 2$.) 
%\begin{equation}
%\end{equation}
$V_3$ is given by
\begin{equation}
V_3 =  (2 \pi)^3 \prod_{i=1}^3 \sqrt{n^{i2} \left(R_1^i\right)^2 + \hat{m}^{i2}
\left(R_2^i\right)^2},
\end{equation} 
where $R_{1,2}^i$ are the  radii of the two dimensions of the $i^{\rm th}$ two-torus, 
$\hat{m}^i = m^i$ for $i=1,2$, $\hat{m}^3 = \tilde{m}^3$,
and the wrapping numbers $(n^i,\hat{m}^i)$ are given in 
Table~\ref{cycles3family}.
The total internal volume is given by
\begin{equation}
V_6 = \frac{ (2 \pi)^6}{4} \prod_{i=1}^3 R_1^i R_2^i.
\end{equation}
The factor of 1/4 comes from the fact that we are orbifolding $T^6$ by
${\bf Z}_2 \times {\bf Z}_2$, which is an Abelian group of order 4.
The Planck scale and Yang-Mills couplings are related to the string
coupling $g_s$ by
\begin{equation}
(M_P^{(4d)})^2 = \frac{ M_s^8 V_6 }{ (2 \pi)^7 g_s^2}
\label{mpms}
\end{equation}
and
\begin{equation}
\frac{1}{g_{YM}^2} = \frac{ M_s^3 V_3}{(2 \pi)^4 g_s}.
\end{equation}

In terms of the complex structure moduli $\chi_i = R_2^i/R_1^i$,
\begin{equation}
g_{YM}^2 = \frac{\sqrt{2 \pi}M_s}{2 M_P^{(4d)}} \frac{\sqrt{\chi_1 \chi_2 \chi_3}}{
\prod_{i=1}^3 \sqrt{n^{i2} + \hat{m}^{i2} \chi_i^2}}
\end{equation}
Supersymmetry requires that $\chi_1 : \chi_2 : \chi_3 = 1:3:2$. Therefore,
\begin{equation}\label{coupling}
g_{YM}^2 = \frac{\sqrt{12 \pi} M_s}{2 M_P^{(4d)}}.
\frac{\chi^{3/2}}{\sqrt{[(n^1)^2 + (m^1)^2 \chi^2][
(n^2)^2 + 9 (m^2)^2 \chi^2][(n^3)^2 + 4 (\tilde{m}^3)^2 \chi^2]}},
\end{equation}
where $\chi \equiv \chi_1$.

At a scale $M$ below the string scale, the coupling $\alpha_a = g_a^2/4\pi$
of the $a^{\rm th}$ gauge factor is given (at one loop) by
\beq
\frac{1}{\alpha_a(M)} = \frac{c_a(\chi)}{\alpha_G(\chi)} + b_a t, \eeq
where 
\beq 
\alpha_G(\chi) = \sqrt{\frac{{3}}{{\pi}}} \frac{ M_s}{4  M_P^{(4d)}} \chi^{3/2}
\eeq
and 
\beq 
t = \frac{1}{2\pi} \ln \frac{M_s}{M}.
\eeq
For $M =M_Z$ and $M_s \sim  M_P^{(4d)}$ one has $t \sim 6 $.
The $c_a$ and $b_a$ are displayed in Table~\ref{gaugefactors}.
For the $b_a$ the contributions $b_a({\rm int})$ from the states at
the intersecting branes (used in the analysis) and those
$b_a({aa})$ from the unwanted $aa$  states
are displayed separately.
The hypercharge $U(1)$ is in general a linear combination
of $U(1)$'s from branes wrapping around different three-cycles:
\begin{equation}
Q_Y = \sum_a d_a Q_a. 
\end{equation}
Its coupling $\alpha_Y$
(or the related $\alpha_1 = \sqrt{5/3} \alpha_Y$ conventionally
used in the MSSM) is not independent of the three other $U(1)$'s,
but is included for convenience. It is given by the relation
\begin{equation}
\frac{1}{\alpha_Y} = \sum_a d_a^2 \frac{1}{\alpha_a}.
\end{equation}
\begin{table} \footnotesize
%[htb] \footnotesize
\renewcommand{\arraystretch}{1.25}
\begin{center}
\begin{tabular}{|c||c|c|c|}
Group $a$& $c_a$ &
$b_a({\rm int})$ & $b_a({aa})$
\\
\hline
$SU(3)$  &  $1+\chi^2$  &  $-1$ & 9 \\
$SU(2)$  &  $1 + 9 \chi^2$  & 18  &  6  \\
$B-L$    &  $\frac{10}{9} \left( 1 + \chi^2 \right)$  &  $\frac{64}{3}$  &
   $-$ \\
$Q_8 \pm Q_{8'}$  &  $12 \chi^2 $  &  80  &  144  \\
\hline
$\frac{1}{\alpha_1} = \frac{3}{5} \frac{1}{\alpha_Y}$ & 
  $\frac 16 + \frac{59\chi^2}{30}$  &  $\frac{76}{5}$  &  $\frac{108}{5}$ \\
\hline
$Sp(2)_B$  & $6\chi^2$  &  $-4$  &  $-$  \\
$Sp(2)_A$  & 2  &  $-6$  &  $-$  \\
$Sp(4) $  & $2 \chi^2$  &  $-5$  &  3  \\
\end{tabular}
\end{center}
\caption{\small Coefficients $c_a$ of $1/\alpha_G$, and
$\beta$ functions $b_a$ for the $a^{\rm th}$ gauge group.
The contributions $b_a({\rm int})$ from the intersecting
branes (used in the analysis), and the additional contributions
$b_a({aa})$ from unwanted $aa$ states not localized at the
intersections are both displayed.  $Sp(2)_B$ and
$Sp(2)_A$ refer to the $Sp(2)$ groups associated with
the $B2$ and $A2$ branes, respectively. The gauge coupling for
hypercharge is not independent of that for
$B-L$ and $Q_8 \pm Q_{8'}$, but is shown for completeness.
For comparison, the $\beta$ functions $b_a({\rm MSSM})$
in the MSSM for $SU(3)$, $SU(2)$, and $1/\alpha_1$ are
respectively $-3$, 1, and $33/5$.
\label{gaugefactors}}
\end{table} 

\begin{figure}
\label{ewgauge}
\centering
\begin{minipage}[c]{0.33\textwidth}
\centering
 \epsfysize=7cm\epsfbox{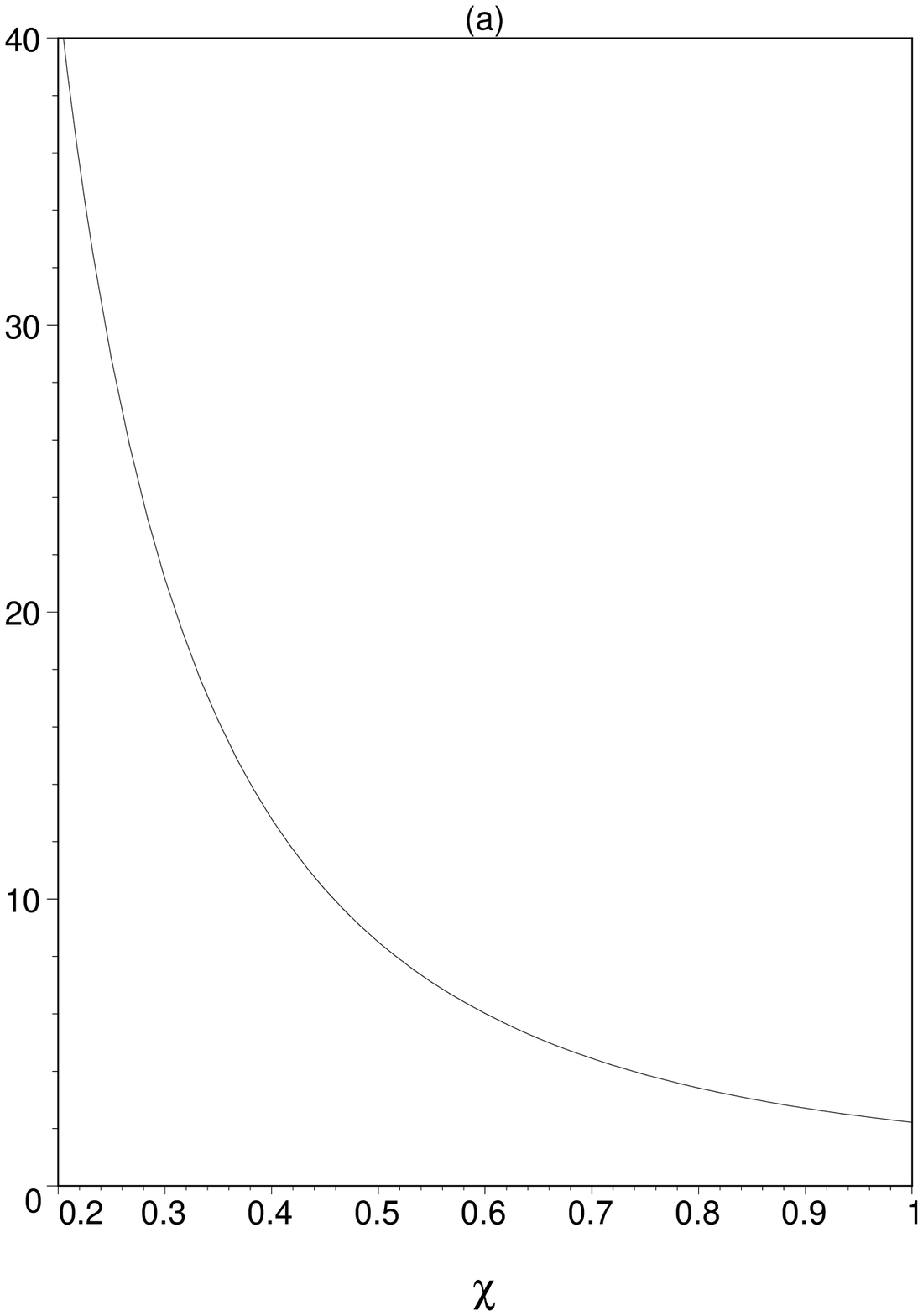}
 \end{minipage}%
\begin{minipage}[c]{0.33\textwidth}
\centering
 \epsfysize=7cm\epsfbox{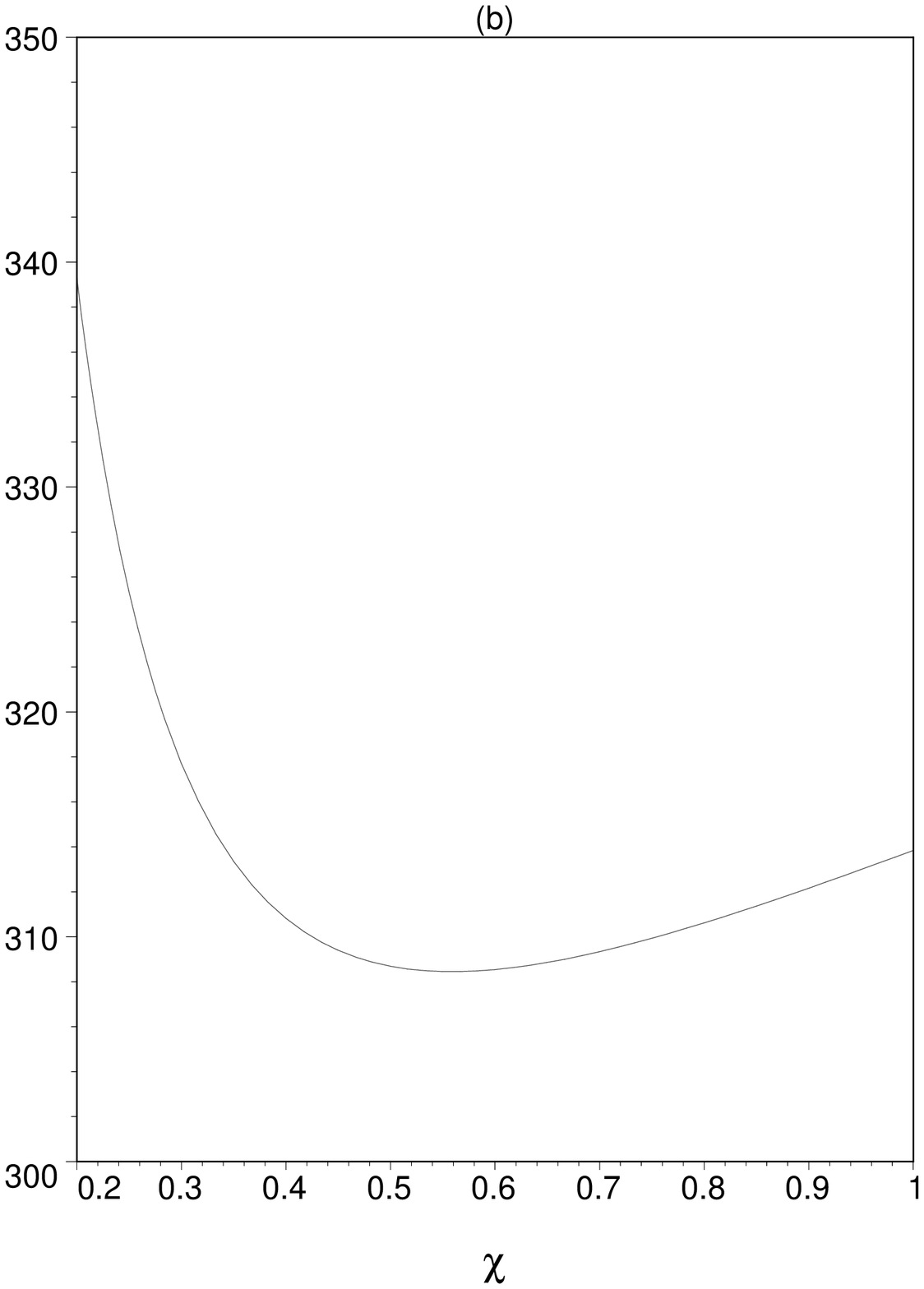}
 \end{minipage}%
\begin{minipage}[c]{0.33\textwidth}
\centering
 \epsfysize=7cm\epsfbox{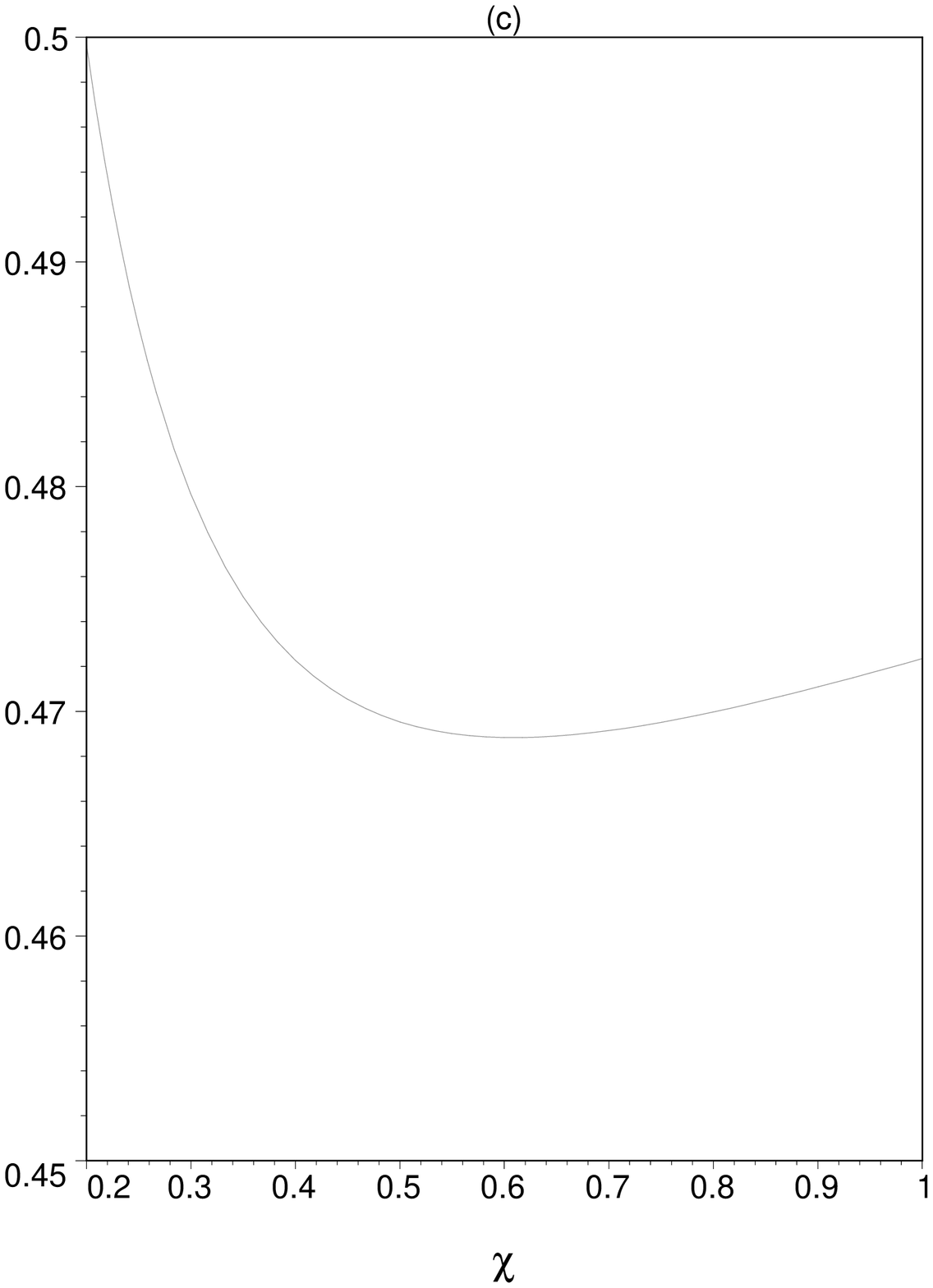}
 \end{minipage}%
\caption{Predicted values of (a) $1/\alpha_3, (b) 1/\alpha = 1/\alpha_Y +
1/\alpha_2$, and (c)
the weak angle $\sin^2 \theta_W = \alpha_Y/(\alpha_2 + \alpha_Y)$
at the electroweak scale as a function of $\chi $ for
$M_P^{(4d)}  = M_s$.
Only the contributions
 $b_a({\rm int})$ of the states localized at the brane
 intersections (as well as the gauge bosons) to the running are
 included. The experimental values are $\sim$ 8.5, 128,
and 0.23, respectively.}
\end{figure}

From (\ref{mpms}), along with the requirement $M_s^6 V_6 \simgr (2 \pi)^6$
(i.e., the compactification radii cannot be smaller than the
string scale) and $g_s^2 \simle 4 \pi$ (perturbative string theory),
we expect that $M_P^{(4d)} \simgr M_s/\sqrt{8}\pi$.
We will generally take  $M_P^{(4d)} = M_s$ in our numerical examples.
Near unification of, e.g., the $SU(3)$ and $SU(2)$ couplings
at $M_s$ prefers small $\chi$, but the overall scale $\alpha_G$
requires that $\chi$ cannot be too small.  The best results
are for $\chi \sim 0.5$.
The predicted MSSM gauge couplings at the electroweak scale
are presented
as a function of $\chi$ in Figure~\ref{ewgauge}.
It is seen that
the predicted value of $\alpha_3$ is quite close to the experimental
value for $\chi \sim 0.5$ ($1/\alpha_3 \sim 8.5$). 
However,
$\alpha$ is predicted
to be much  too small, mainly because of the contributions of the exotic
states to the running,
while  $\sin^2 \theta_W$ is predicted too large by a factor $\sim 2$. 
(The predicted value of $\sin^2 \theta_W$ at  the string scale $M_s$
ranges from 0.78 to
0.73 for $\chi$ varying from 0 to $\infty$, while  $1/\alpha_G \sim 12$
for $\chi \sim 0.5$ and $M_P^{(4d)}  = M_s$.)

\begin{figure}
\label{spcouplings}
\centering
\begin{minipage}[c]{0.45\textwidth}
\centering
 \epsfysize=7cm\epsfbox{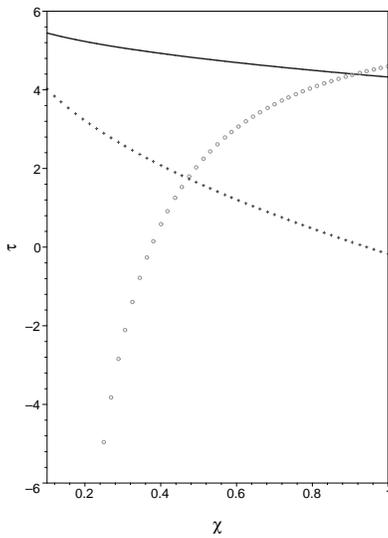}
 \end{minipage}%
\parbox[c]{0.05\textwidth}{}
\begin{minipage}[c]{0.5\textwidth}
\centering
\caption{Scales $\tau = \frac{1}{2\pi} \ln \frac{M}{M_Z}$,
where $M$ is the scale at
which the $Sp$ group becomes strongly coupled for
$M_P^{(4d)}  = M_s$. The curves are $Sp(2)_A$ (solid),
$Sp(2)_B$ (circles), $Sp(4)$ (crosses).}
\end{minipage}
\end{figure}

Although the predictions for the MSSM gauge couplings are unrealistic,
the quasi-hidden sector $Sp$ groups are all asymptotically free. 
Figure~\ref{spcouplings} displays the scales at which each $Sp$
group becomes strongly coupled as a function of $\chi$.
All three groups 
become strong above the electroweak scale for
$M_P^{(4d)}  = M_s$ and
$\chi \simgr 0.4$.
For $M_P^{(4d)}  < M_s$  the couplings become strong at higher scales
(e.g., $Sp(2)_B$ would become strong at around $10^{15}$ GeV for
$M_P^{(4d)}  = M_s/3$  and $\chi = 0.5$, with possible implications
for neutrino mass, as mentioned in Section~\ref{additional}).
The implications will be further discussed in Section~\ref{strongly}
and in~\cite{strongsector}.

\section{Implications of a Strongly Coupled Sector}
\label{strongly}
We have seen that for sufficiently low values of  $M_P^{(4d)}  / M_s$
the quasi-hidden sector groups  $Sp(2)_B$,  $Sp(2)_A$, and $Sp(4) $
will become strongly coupled above the electroweak scale. 
This is likely to lead to supersymmetry breaking in the hidden sector,
which can be transmitted to the observable sector by supergravity,
as well as dilaton/moduli stabilization~\cite{strongsector}.
Here we will focus on another aspect, i.e., the confinement of free
$Sp$ charges, expected in analogy with QCD. In particular,
it is plausible to assume that when one of the $Sp(n)$ groups becomes strongly
coupled at some scale $M \gg M_Z$ any states carrying $Sp(n)$ charges become
confined and drop out of the physical spectrum. However, there may be
$Sp$-neutral bound state chiral supermultiplets remaining in the
spectrum\footnote{Supersymmetry breaking associated with a related
gaugino condensation will be transmitted to the ordinary sector only by
weak supergravity effects, and can be ignored for the purposes of the present
discussion.}, which may be required to avoid the introduction of
anomalies for the remaining gauge groups.

$Sp(4)$ is expected to become strong at a high scale. The only state
in the chiral sector with $Sp(4)$ charge is from the $B_1 C_2$
intersection. We denote this state as $\Psi$, which
transforms as $(1,2,1,1,4)$ under
$SU(3)\times SU(2)\times Sp(2)_B\times Sp(2)_A\times Sp(4)$.
The strong $Sp(4)$ forces may lead to
a composite chiral supermultiplet $\Psi^2_{AA}$, where the subscript
indicates an antisymmetrization in both the $SU(2)$ and $Sp(4)$ indices.
$\Psi^2_{AA}$ is a total gauge singlet. Whether or not this composite
state is formed, the confinement of $\Psi$ does not lead to any anomalies for
the residual gauge groups. An anomaly is induced in the $Q_2-SU(2)^2$
vertex, where $Q_2$ is associated with an anomalous $U(1)$ with a massive
gauge boson. This can be regarded as a breaking of $Q_2$ and
is presumably harmless, 
analogous to to the breaking of the global axial $U(1)$ symmetry in QCD. 
The four $SU(2)$ doublets contained in  $\Psi$ drop out of the renormalization
group equations for $SU(2)$ at the decoupling scale, lowering
$b_{SU(2)} $ by 2, but this has
 only a minor impact on the discussion in Section~\ref{gauge}.

$Sp(2)_B$ can become strongly coupled anywhere from a few TeV up
to very high scales such as $10^{15}$ GeV, depending on $\chi$ and
$M_P^{(4d)}  / M_s$.
The fractionally charged $B_2C_1$ states are charged under $Sp(2)_B$.
Let us denote them as $\Phi \equiv (3,1,2,1,1)$ and $\Sigma \equiv
(1,1,2,1,1)$. By our assumptions, these will be confined at the
$Sp(2)_B$ scale. The strong $Sp(2)_B$ binding might form the composite
color triplet $\Phi \Sigma = (3,1,1,1,1)$. This has charge $-1/3$, and
would be a candidate for an exotic $SU(2)$-singlet down-type quark, except
that it has lepton number $L=1$. Furthermore, if this were the only
massless composite, anomalies would be induced in the $Q_Y^3$ and
$Q_Y-SU(3)^2$ vertices. The anomaly-matching condition suggests that, instead
of $\Phi \Sigma$, there is a more complicated spectrum of massless composites.
The simplest possibility is that the spectrum consists
of 
\beqa
 E_4  &  =  &  \Phi \Sigma \bar U \nonumber \\
 N_4  &  =  &  \Phi \Sigma \bar D \nonumber \\
 U_4  &  =  &  \Phi \Sigma \bar E \nonumber \\
 D_4  &  =  &  \Phi \Sigma \bar N,
 \eeqa
i.e., $\Phi \Sigma$ forms bound states with each member of one of the
four families of $SU(2)$-singlet antiquarks and antileptons. The latter
do not have to drop out of the spectrum,
so the composite $U_4, D_4, E_4$, and
$N_4$ are candidates to be the exotic  ($SU(2)$-singlet)
left-handed partners of the
elementary fourth family of antiparticles. This spectrum
matches all of the gauge anomalies, although it does lead
to a (presumably harmless) $(B+L)-Q_8^2$ anomaly, again similar
to the axial $U(1)$ in QCD.

This binding mechanism seems very plausible from the viewpoint
of anomaly matching. However, it is harder to understand from
the actual forces between the constituents, since the
$\bar U, \bar D, \bar E, \bar N$ are not charged under the strong
$Sp(2)_B$ group. (They do carry other gauge charges.)

The decoupling of $\Phi$ and $\Sigma$ and
the appearance of the composite states lead to a net increase of 
$6/5$ for the $\beta$ function for $1/\alpha_1$ at the decoupling scale
($b_{SU(3)}$ and $b_{SU(2)}$ do not change),
but this is a small effect for the specific
numerical example we have displayed. 
For example, for
$M_P^{(4d)}  = M_s$ and $\chi=0.5$, $Sp(4)$ and $Sp(2)_B$
become strong at $\sim 10^{15}$ GeV and $2 \times 10^{6}$ GeV, respectively.
Including both decouplings, the predicted values of $\alpha^{-1}$
and $\sin^2 \theta_W$ decrease by $\sim 6$ and $0.02$, respectively,
compared to those in Figure~\ref{ewgauge}.

$Sp(2)_A$ may also become strongly coupled. However, there are
no chiral states with $Sp(2)_A$ charges.

\section{Discussion}
\label{discussion}
In this paper we have described the phenomenological implications of
a semi-realistic supersymmetric three family model derived from
an orientifold construction. In addition to the MSSM, the model
involves an extended gauge structure, including two additional
\upr \ factors, one of which has family non-universal couplings.
There is also a quasi-hidden sector non-abelian group, which becomes
strongly coupled above the electroweak scale. There are many exotic
chiral supermultiplets, including an exotic ($SU(2)$-singlet) fourth
family of quarks and leptons in which the left-chiral states have
unphysical fractional electric charges. These are presumably confined by the
strong hidden sector interactions, while anomaly constraints imply composite
left-chiral states with the correct charges. The right-chiral states are
elementary. The Yukawa sector~\cite{yukawa} and other aspects
of the hidden sector~\cite{strongsector} will be presented
separately.

As emphasized in the Introduction,  none of the models that
have been constructed are
fully realistic, and it is difficult to know whether
the specific features of a given model are hints of possible
new TeV scale physics, or merely artifacts of the construction.
For that reason, it is useful to contrast some of the features
of this orientifold construction with those of a specific
heterotic model described in~\cite{chl5}. For convenience,
those predictions are described in more detail in the Appendix.

Both models predict additional \upr \ gauge symmetries,
some of which have family-nonuniversal and therefore flavor
changing neutral current couplings. Both are most likely
broken at the TeV scale, but have a possibility of being
broken at an intermediate scale along a $D$-flat direction.
Both also have quasi-hidden nonabelian gauge sectors. This means
that while most of the states in the model are charged under
one or neither of the gauge sectors, there are a few
states which couple to both. (The \upr \ also connect the two sectors.)
These mixed states have fractional charges like $1/6$ or $\pm 1/2$.
A hidden sector is an ideal candidate for dynamical supersymmetry breaking
if it is strongly coupled. In the heterotic example the hidden
sector groups are not asymptotically free. However, in the
orientifold example, the groups are asymptotically free, and may lead to gaugino
condensation, dilaton/moduli stabilization, and charge  confinement,
modifying the low energy spectrum.

Both models involve exotic states, often with no satisfactory
means of generating fermion masses. These include additional Higgs
doublets and singlets, suggesting such effects as a rich spectrum
of Higgs particles, neutralinos, and charginos, perhaps with 
nonstandard couplings due to mixing and flavor changing effects.
The effective $\mu$ terms are either missing or nonstandard.
There may also be vector pairs of additional quarks and leptons.
In the orientifold case, the left-chiral states are composite
and their right-chiral partners elementary. The orientifold
model also contains unwanted adjoints. There may be
mixing between lepton and Higgs doublets, leading to
lepton number violation. This was required in the heterotic
case (where baryon number violation was also possible for one
flat direction), and optional for the orientifold.

Although the Yukawa couplings 
have tree-level contributions in string
perturbation theory (in orders of $g_s$) in the two constructions,
they have different origins from the worldsheet perspective.
In the orientifold model, the Yukawa couplings arise from non-perturbative
effects (worldsheet instantons) in the worldsheet conformal field theory. 
In the CHL5 model, the Yukawa
couplings are tree-level from the worldsheet point of view.
We note, however, that in some heterotic string constructions, the
quarks, leptons and the Higgs fields are localized at different orbifold
fixed points (see e.g., \cite{cvet} and references therein.).
 In these constructions, the Yukawa couplings also
come from worldsheet instantons.
Both constructions can yield masses and mixings for some, but
not all, of the fermions, but the details depend on the mechanism
of supersymmetry breaking. Neither has an obvious mechanism for
a neutrino seesaw, except possibly for the case of an intermediate
scale breaking of a \upr.

In the heterotic model the gauge unification predictions are
non-standard (and not very successful) due to the combination
of exotic particles contributing to the running of the gauge
couplings and higher Ka\v c-Moody levels at the string scale.
The orientifold predictions are also non-standard: the gauge
groups are associated with different branes, and have
non-standard moduli-dependent boundary conditions at the string
scale, and there are also exotic particles contributing to the
running, leading to electroweak couplings that are too small.
Resolutions of these difficulties
in more successful constructions might
involve avoiding these non-standard features, having exotics
which fall into complete grand unification multiplets,
or invoking cancellations of effects occurring by accident
or due to some other mechanism.

\begin{acknowledgments}
We are grateful to Angel Uranga and Jing Wang
for useful discussions and collaborations on related work.
This work was supported by
the DOE grants  EY-76-02-3071
and  DE-FG02-95ER40896;
by the National Science Foundation Grant No. PHY99-07949;
 by the University of Pennsylvania School of Arts and Sciences
Dean's fund (MC and GS); by the University of Wisconsin at Madison (PL);
by the W. M.
Keck Foundation as a Keck Distinguished Visiting Professor at the
Institute for Advanced Study (PL);
and by the ITP, Santa Barbara,  the ITP workshop
on Brane Worlds and the  Newton Institute, Cambridge. 
\end{acknowledgments}

\appendix
\section*{}

As described in the Introduction, there has been considerable
study of semi-realistic perturbative heterotic string constructions,
including
a class of free-fermionic string models which contain
the gauge group and matter content of the MSSM~\cite{fny,CHL}.
Such constructions generally involve additional gauge factors and many extra
matter supermultiplets. However, they also contain an anomalous $U(1)_A$ symmetry
and a corresponding Fayet-Iliopoulos contribution to the $U(1)_A$ $D$-term.
Maintaining supersymmetry at the string scale requires that some of the
fields in the effective four-dimensional theory must acquire compensating
vacuum expectation values (VEVs) near the string scale, while maintaining
$F$-flatness and $D$-flatness for the other gauge factors. These break some of the
extra gauge symmetries, remove some of the apparently massless states from
the low energy effective theory, and require that the theory be restabilized, i.e.,
the superpotential for the remaining massless states must be recalculated when some of
the fields are replaced by their string-scale VEVs. A systematic procedure was
developed in~\cite{flat} to classify the flat directions associated with
non-abelian singlet fields.
This was used in~\cite{chl5} to investigate
the flat directions and related low energy phenomenology 
in detail for a promising model due
originally to Chaudhuri, Hockney, and Lykken~\cite{CHL} (CHL5).
Flat directions associated with non-singlets were studied in~\cite{clew}.
The procedures were used to study the flat directions in
a class of models due
to Faraggi, Nanopoulos, and Yuan~\cite{fny} in~\cite{TAM}.

The features of perturbative heterotic string models  are illustrated by 
the prototypical
example of the CHL5 model. These include:

\begin{itemize}
\item One or more additional (nonanomalous) $U(1)'$ gauge factors. The
associated $Z'$ gauge boson is typically expected to be lighter than around
1 TeV~\cite{ewscale}, although for one flat direction studied the breaking could
be at an intermediate scale~\cite{intscale}. The $Z'$ couplings are family
nonuniversal, leading to flavor changing neutral currents (FCNC)~\cite{fcnc}.

\item Additional non-abelian gauge factors. These could, in principle, 
play a role in dynamical supersymmetry breaking. However, in the model studied
the factors do not become strongly coupled below the string scale (i.e, they
are not asymptotically free). These extra gauge factors are quasi-hidden, i.e.,
most matter multiplets transform nontrivially under the standard model
$SU(3) \times SU(2)$ group or under the hidden sector group (or neither),
but not both. However, there are a few exceptions which are charged
under both. The extra $U(1)'$ typically couple to both sectors.

\item
There are many exotic supermultiplets in the model, including an extra $d$-type quark,
extra Higgs/lepton doublets, and many non-abelian singlets. For many, there is no
mechanism to give a significant mass to the fermions. This is a major flaw
of most such models. (One exception, which however has unrealistic
Yukawa couplings, is described
in~\cite{TAM}.) The spectrum includes a number of charge $\pm 1/2$ states. These
are all charged under the quasi-hidden sector group, and potentially could
disappear from the spectrum if the hidden sector charges are
confined~\cite{crypton}. However, as noted above, the hidden sector factors in the
CHL5 model are not strongly coupled, so this mechanism does not occur.

\item
There are more than the two Higgs doublets of the MSSM.
In addition to the more complicated Higgs spectrum,
there is a possibility of Higgs mediated FCNC.
Models with an electroweak scale $U(1)'$ can generically provide a natural solution to
the  $\mu$ problem~\cite{SY,ewscale}, which is generated dynamically by the VEV of the
field that breaks the $U(1)'$. However, in the CHL5 model the effective $\mu$
terms are non-canonical,
connecting Higgs doublets which generate the $t$ and $b$ masses to others.
In the specific examples studied, one of the needed  effective $\mu$ terms is
absent, leading to an unwanted global symmetry.  

\item
The model has gauge coupling unification. However, the detailed predictions for
the low energy couplings differ from the MSSM (and from experiment) because of
the additional matter fields as well as higher Ka\v c-Moody levels for the $U(1)$
factors.

\item
The Yukawa couplings at the string scale are either $g$, $g/\sqrt{2}$, or 0, 
where $g=
O(1)$ is the gauge coupling, allowing large masses for the $t$ and $b$, and the
possibility of radiative electroweak breaking. Smaller Yukawa couplings can be
associated with
higher dimensional operators that become cubic after vacuum restabilization.
CHL5 contains $t-b$ and an unphysical $\mu-\tau$ universality, a noncanonical
$b-\tau$ relation, and a nontrivial (but unphysical) $d$ quark texture for
one flat direction.

\item
The models violate $R$-parity and lepton number (due to mixing between
lepton and Higgs doublets), so there is no
stable neutralino. Baryon number is violated for one flat direction,
leading to proton decay and $n-\bar{n}$ oscillations (with rates that cannot be
calculated without resolving the problem of the massless exotics). 

\item
There is no obvious mechanism for a neutrino seesaw except for the cases
in which one of the $U(1)'$ gauge factors is broken at an intermediate
scale~\cite{intscale}.

\item
When phenomenological soft supersymmetry breaking parameters are introduced by
hand at the string scale, one can calculate the symmetry breaking and the
spectrum of the Higgs and supersymmetry particles. One can obtain a sufficiently
large $Z'$ mass (e.g., $>$ 700 GeV) and small $Z-Z'$ mixing (e.g., $<$ 0.005)
for somewhat tuned values of the soft parameters. The large $Z'$ mass
scale is set by the soft breaking parameters, implying a
spectrum quite different from most of the parameter space of the MSSM: typically,
the squark and slepton masses are in the TeV range (except possibly for the third
family), but there is a richer spectrum of Higgs particles, charginos, and
neutralinos~\cite{nonstandard}.

\end{itemize}

\end{document}